\documentclass[aps,prl,showpacs,amssymb, nobibnotes,twocolumn]{revtex4-1}

\usepackage[dvips]{epsfig}
\usepackage{graphics,graphicx}

\begin{document}

\title{Experimental demonstration of three-photon Coherent Population Trapping in an ion cloud}
\author{M. Collombon}
\author{C. Chatou}
\author{G. Hagel}
\author{J. Pedregosa-Gutierrez}
\author{M. Houssin}
\author{M. Knoop}
\author{C. Champenois}
\email{caroline.champenois@univ-amu.fr}

\affiliation{Aix Marseille Univ, CNRS, PIIM,  Marseille, France}

\date{\today}

\begin{abstract}
A novel protocol of  interrogation based on coherent population trapping in an $N$-level scheme atomic system leads to dark resonances involving three different  photons. An ensemble of several hundreds of radio-frequency trapped Ca$^+$ ions is probed by three lasers simultaneously locked onto the same optical frequency comb, resulting in high-contrast spectral lines referenced to an atomic transition in the THz domain.  We discuss the cause of uncertainties and limitations for this method and show that reaching a sub-kHz resolution is experimentally accessible via this interrogation protocol.
 \end{abstract}


\pacs{   }

\maketitle

The quantum interferences between two excitation paths which are responsible for coherent population trapping (CPT) are an example of a quantum effect based on atomic coherences \cite{arimondo96}. When this interference occurs in a $\Lambda$-scheme atomic system,  the atomic population is trapped in a coherent superposition of the two ground sub-states, dressed by the coupling photons. If the two ground dressed sub-states are stable,  the excited state is no more populated. This population trapping can  be observed by a dark resonance in the fluorescence signal \cite{alzetta76} or by the cancellation of the laser absorption \cite{schmidt96}. Two-photon CPT has proven its relevance as a resource for high precision measurement in magnetometry \cite{nagel98,schwindt04} and in the so-called "CPT-microwave clock" where no microwave cavity is needed anymore \cite{cyr93,wynands99,vanier05}. The best reported performances concerning short term stability are reached with vapour cell clock with fractional frequency stability of   few $10^{-13} /\sqrt{\tau}$ \cite{yun17,hafiz17}. Cold atom clocks are expected to perform better regarding long term stability and an instability of the order of  $3\times 10^{-13}$  after one hour of averaging time is reported in \cite{liu17}. In this paper, we report the observation of a three-photon CPT in a cold cloud of trapped ions and discuss its main causes of shift and broadening.

 In CPT-microwave clocks, two phase-coherent lasers  perform optical spectroscopy of the  GHz-transition between hyperfine sub-levels of the heaviest alkalies Rb and Cs. When the laser difference frequency matches the ground-state splitting, the atomic population is trapped in a dark state and the optical signal is used to reference  this frequency difference to the GHz-transition. Sub-Doppler spectroscopy is reached by exploiting the Lamb-Dicke effect \cite{dicke53} which provides a first-order Doppler cancellation whenever the displacement of the absorbers over successive excitation is smaller than $\lambda/(2\pi)$ with $\lambda$ the transition wavelength, of the order of centimeters for microwave transition and micrometers for optical transitions. Experimental results  in room temperature cells where the atom mean-free-path is reduced to millimeter scale by  filling with buffer gas or by scaling the cell size to millimeter prove that the Lamb-Dicke regime can be reached for the microwave transition even if it is excited by means of two lasers operating in the optical range. The contrast of the dark line is then limited by the relaxation of the coherence between the two sub-states. It can be induced by the collisions of the atoms with the buffer gas and/or the cell glass \cite{kitching02} and by the noise on the relative phase between the optical fields \cite{hemmer89,liu17}.

We investigated in \cite{champenois06} a three-photon CPT which occurs in a four-level atomic system showing a $N$-shaped laser interaction scheme, where three out of the four involved states are stable or metastable. Extending from two to three  lasers involved in the dark resonance condition allows the cancellation of the first order Doppler effect by a geometric phase matching of the three laser wave-vectors \cite{grynberg76,hong05,champenois07,ryabtsev11,barker16}, simulating a Lamb-Dicke effect with an effective infinite wavelength. This example of $N$-level scheme can be found  in the heaviest alkaline-earth ions Ca$^+$, Sr$^+$ and Ba$^+$\cite{champenois06} and in alkaline-earth-metal neutral atoms like Sr and Yb \cite{barker16}. We report here the first observation of a three-photon dark resonance in a cloud of Ca$^+$ ions, stored in a linear quadrupole trap, and laser-cooled by  Doppler laser cooling. In this ion, the three  optical fields required to build the coherent dark state lie in the optical and near infra-red domain, spanning more than one octave. Therefore, their phase coherence is insured by an ultra-stable laser through a simultaneous lock on an optical frequency comb (OFC) \cite{scharnhorst15,collombon19}. The dark resonance condition defines  a combination of the three optical frequencies, which is referenced to a magnetic dipole transition at 1.82~THz. 

In the following, we first present the experimental conditions for observing a three-photon dark resonance in the fluorescence of  a cloud of trapped ions. We review the major effects which  contribute to the  linewidth, frequency shift and contrast of the dark line as there are the Doppler effect, the Zeeman effect and  power-induced effects.

\section{Condition for observations of a three-photon CPT}

\begin{figure*}[htbp]
\begin{center}
\includegraphics[width=12cm]{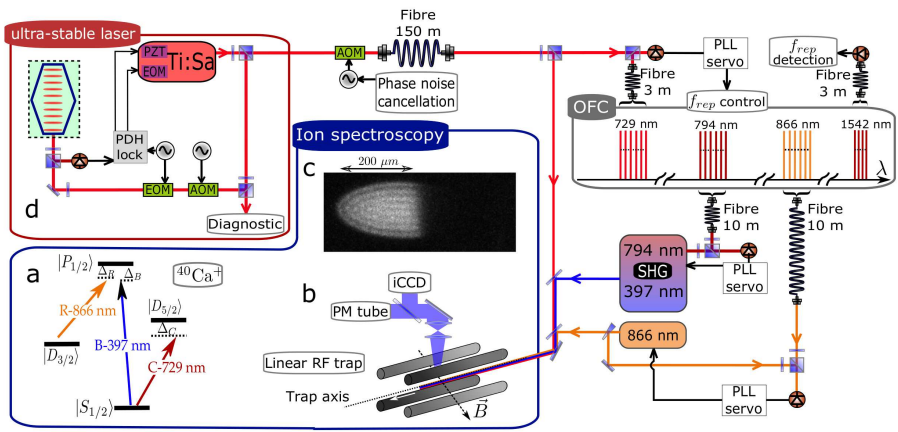}
\caption{{\bf a} : Transition scheme for the 3-photon CPT in Ca$^+$, {\bf b} :  Schematic of the experimental set-up : the three laser beams are propagating along the trap symmetry axis, the linear radio-frequency trap is represented by its four rods, the laser induced fluorescence is recorded on a photomultiplier (PM tube) for photon counting and a intensified CCD camera for spatial resolution of the fluorescence. {\bf c} : picture of a cloud made of $710~(\pm 35)$ ions, the ions shelved in the metastable D$_{5/2}$ state or trapped in the coherent superposition are dark and split from the bright ions because they do not feel the radiation pressure induced by the cooling laser on the bright ions. {\bf d}: set-up for the laser phase-lock, based on an OFC locked onto an ultra-stable Ti:Sa laser. AOM : acousto-optic modulator, EOM : electro-optic modulator, PDH : Pound-Drever-Hall, PLL : phase-lock loop, SHG : second harmonic generation.}
\label{fig_demo3P}
\end{center}
\end{figure*}

The three-photon dark line is observed in the laser-induced fluorescence emitted by a cloud of $^{40}$Ca$^+$ ions,  stored in a linear RF quadrupole trap. The calcium ions are Doppler laser-cooled on their  resonance  transition 4S$_{1/2}\to4$P$_{1/2}$ at 396.85~nm (label $B$).  This transition is not closed and once in the excited state 4P$_{1/2}$, the ions can decay to the metastable state 3D$_{3/2}$ with a probability $\beta=0.064$ \cite{ramm13}. Keeping the ions within the cooling cycle thus implies a second "repumping" laser, tuned to the dipole transition 3D$_{3/2}\to4$P$_{1/2}$  at 866.21~nm (label $R$). The third laser involved in the CPT process is resonant with the electric quadrupole transition 4S$_{1/2}\to3$D$_{5/2}$ at 729.15~nm (label $C$, see Fig.~\ref{fig_demo3P}, {\bf a} for the transition scheme). 

Because it is based on the second order expansion of the interaction Hamiltonian, this last transition has a typical coupling strength which is 8 orders of magnitude smaller than the one induced on the dipole transition involved in the laser cooling \cite{knoop04}. Despite of the weakness of the laser-atom interaction on this transition, it can play a major role in the internal state dynamics provided that a  resonance condition is fulfilled \cite{champenois06}. By a partial diagonalisation of the system, this   condition can be extrapolated from two-photon $\Lambda$-scheme dark resonances \cite{arimondo96} and it  writes 
\begin{equation}
\Delta_R=\Delta_B-\Delta_C-\delta_C
\label{accord_D}
\end{equation}
 where $\Delta_R, \Delta_B, \Delta_C$ are the one-photon detunings of the three lasers and $\delta_C$ is the light-shift induced by the quadrupole coupling on the 729~nm transition \cite{champenois06} (see  appendix \ref{app}). The trapping  state is a coherent superposition of the three stable and metastable dressed states   that is not coupled by laser excitation and once trapped in this state, the ions do not emit any photon.
  
  When fulfilled, the three-photon resonance condition implies a strong relation between the three laser frequencies
\begin{equation}
\omega_R+\omega_C-\omega_B+\delta_C=\omega_{THz}
\label{accord}
\end{equation}
with  $\omega_{THz}$  the frequency of the magnetic dipole transition between  3D$_{3/2}$ and 3D$_{5/2}$, which appears as the  reference transition. The frequency $\omega_{THz}$ is  1.82~THz  in Ca$^+$ and its absolute value  is known with a $\pm 8$~Hz uncertainty through  Raman spectroscopy on a single trapped ion \cite{yamazaki08b,solaro18}.   Considering the typical intensity and  detuning for the laser at 729~nm,  $\delta_C$ is below 100~Hz and can be neglected  in the results reported in the following.

In the experiments presented here, the three lasers co-propagate along the symmetry axis of a linear quadrupole RF-trap (see Fig.~\ref{fig_demo3P},{\bf b}) and the effective wave-vector ${\mathbf \Delta\vec{ k}}$ controlling the first order Doppler effect on the dark line is the one of the magnetic dipole transition $k_{THz}=2\pi/\lambda_{THz}$  with $\lambda_{THz}$ the 3D$_{3/2}$-3D$_{5/2}$ transition wavelength, equal to 165~$\mu$m. The quadrupole trap is described in \cite{champenois13, kamsap15t},   its main characteristics are an inner radius of 3.93~mm for a rod radius of 4.5~mm \cite{pedregosa10} and an RF trapping frequency of 5.2~MHz. For the work presented here, it is operated with an RF-voltage difference between neighboring rods of 826~V$_{pp}$ (Mathieu parameter $q_x=0.24$) and a typical cloud contains a few tens to  thousands ions. The Doppler-laser cooling drives the ion cloud  from a gas, through the liquid,  to a crystal phase \cite{drewsen98} with a temperature estimated to be of the order of 10~mK. Once in the liquid and crystal phase, the ion cloud forms an ellipsoid \cite{turner87,hornekaer02}   with a diameter ranging from 80 to 280~$\mu$m and a length ranging from 120 to 740~$\mu$m  for a  number of ions comprised between 40 and 2750 (see Fig.~~\ref{fig_demo3P},{\bf c}). The 397~nm and 866~nm lasers have an elliptical cross-section, with an aspect ratio of 2 and a mean-squared diameter at the position of the cloud  equal to 4.0~mm and 4.7~mm, respectively. The laser intensity and wave-vectors can  be considered  uniform all over the ion cloud. The 729~nm laser has the smallest size with a waist diameter measured to $300 (\pm 20)~\mu$m. It is still larger than the largest of the cloud diameters but  its intensity is not uniform over the largest clouds. Keeping all the atoms inside the three laser beams cancels any broadening   induced by finite interaction time due to atom motion, which is an identified broadening cause of two-photon CPT lines when observed on an atomic beam or a gas in a cell \cite{thomas81,brandt97,wynands99}. 

 For the observation of the dark lines, the three involved lasers are admitted continuously on the cloud. As well-known for the two-photon CPT, the stability of the phase  relation between the three dressing lasers is mandatory to reach a stationary dark state.  To that purpose, we use a commercial  OFC to transfer the phase stability between the three lasers\cite{collombon19}. We take advantage of an ultra stable laser at 729~nm (relative Allan variance of $10^{-14}$ at 1~s) to serve as a local reference for the offset-free OFC, produced by frequency difference in a non-linear crystal (TOPTICA DFC CORE+). The OFC repetition rate $f_{rep}$ is 80~MHz and three dedicated coherent outputs at 729, 794 and 866~nm allow three simultaneous phase locks (the 397~nm radiation is produced by second harmonic generation). First, the OFC is locked onto the 729~nm laser by a phase-locked loop (PLL) using the beat signal of the ultra stable laser and the closest comb eigen-mode. Then, the  794 and 866~nm lasers are locked onto the OFC by the same technique.  Their frequency  measurement requires to identify the indexes $N_{B,R,C}$ of their closest eigen-mode emitted by the OFC, and to measure the relative value  of the laser frequency compared to this eigen-mode  which is bound by $\pm f_{rep}/2$.  The indices  $N_{B,R,C}$ are determined without ambiguity with the help of a wave-meter  with an accuracy of $\pm 10$~MHz. The uncertainty on each laser frequency measurement lies in the kHz range and is due to $N_{B,R,C}\times \sigma_{rep}$ with $\sigma_{rep}$ the uncertainty of the repetition rate equal on average to 1.5~mHz (one standard deviation).  The resulting uncertainty on the deduced THz frequency benefits from the combination $(N_B-N_R-N_C)\times \sigma_{rep}$ which is of the order of 34~Hz, two orders of magnitude lower than the optical frequency uncertainty.

\begin{figure}[htbp]
\begin{center}
\includegraphics[height=4.2cm]{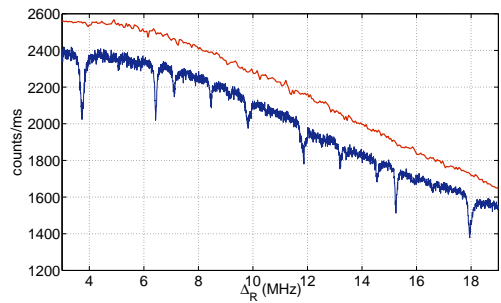}
\caption{Laser induced fluorescence of a cloud of laser cooled 560~$(\pm 25)$ ions versus the detuning of the repumping $R$-laser. The top curve shows photon counts when the cooling and repumping lasers are on ($P_B=10$~mW, $P_R=1$~mW). For the lower curve, the 729~nm laser is also applied ($P_C=12.1$~mW). Background : $570 (\pm 25)$~counts/ms \label{fig_spectre}}
\end{center}
\end{figure}

   Spectra  are observed when collecting the photons emitted at 397~nm on the 4P$_{1/2}\to4$S$_{1/2}$ transition while the frequency of the $R$-laser is scanned.  Typical laser powers are 10 to 20~mW at 397~nm ($P_B$), 0.5 to 5~mW at 866~nm ($P_R$) and  5 to 25~mW at 729~nm ($P_C$).  A complete spectrum width is larger than 100~MHz and figure~\ref{fig_spectre} shows a portion of such a spectrum, selected  around the  three-photon resonance condition as given by Eq.~\ref{accord_D}.  By comparing the two plots of Fig.~\ref{fig_spectre}, showing the fluorescence signal with and without the $C$-laser exciting the weak transition, one can deduce that some ions are "shelved" in the metastable D$_{5/2}$ state \cite{nagourney86}, independently of any resonant condition. The narrow  features superimposed to the reduced signal (blue line) are the signature of the three-photon CPT. They correspond to the trapping of a fraction of the  ions from the cooling cycle to the dark state. With the chosen detection scheme, it is not possible to quantify the number of ions transferred  from the metastable state to the dark state. More than half of the ions remain bright and thus are laser-cooled. By sympathetic cooling of the dark atoms \cite{larson86}, the ion cloud remains in a liquid phase throughout  the complete frequency scan, even if the radiation pressure is responsible for a spatial separation of the bright and dark ions (see Fig.~\ref{fig_demo3P},{\bf c}). This sympathetic cooling  offers the great advantage of keeping constant the number of  ions inside the laser beam  during the whole recording. 

The splitting of the dark line into several pairs of lines is due to the local magnetic field which lifts the degeneracy of the Zeeman sub-states. A magnetic field of the order of 1 Gauss is applied and the three laser polarizations are linear, perpendicular to the trap axis and nearly perpendicular to the local magnetic field. We label each transition by a number $m_{THz}$ according to its Zeeman shift on the THz transition frequency $\delta_Z (m_{THz})=m_{THz} \mu_B B$. With $g_{3/2}$ and $g_{5/2}$,  the Land\'e factors of the D$_{3/2}$ and D$_{5/2}$ states, and $m_J (D_{3/2})$ and $m_J (D_{5/2})$ the Zeeman substates involved in the transition, $m_{THz}=g_{5/2}m_J (D_{5/2})-g_{3/2}m_J (D_{3/2})$.  Table \ref{tab_mTHz} gathers the transitions based on  the largest  couplings (quantified by their relative Rabi frequencies) imposed by the chosen laser polarization, relative to the local magnetic field and controlled by the selection rules of dipole and quadrupole transitions \cite{sobelman}.
 \begin{table}[htb]
\caption{Zeeman sub-states of the two metastable states giving rise to the observed THz transitions. The corresponding THz transitions are labelled by the number $m_{THz}$, defined by the Zeeman shift  $\delta_Z (m_{THz})=m_{THz} \mu_B B$ (for the labelling only, $m_{THz}$ is rounded in the assumption of a Land{\'e} $g$-factor equal to 2 for the electron). ${\overline\Omega_C}$ and ${\overline\Omega_R}$ are the Rabi frequencies relative to  the one of $m_{THz}=\pm 21/5$. All the transitions share the same Rabi frequency on the $B$-transition. }
\begin{center}
\begin{tabular}{|c|c|c|c|c|c|}
\hline
$m_J (D_{3/2})$ & $m_J (D_{5/2})$ & $m_{THz}$ & ${\overline\Omega_C}$ & ${\overline\Omega_R}$ \\
\hline
 $\mp 3/2$  & $ \pm 5/2 $ & $ \pm 21/5$ & 1 & 1 \\
$\pm1/2$  & $\pm5/2 $ & $\pm 13/5  $ & 1 & $1/\sqrt{3}$\\
$\mp1/2 $& $ \pm 3/2$ & $\pm 11/5$ & $1/\sqrt{5}$  & $1/\sqrt{3}$\\
$ \pm1/2$ & $ \pm 3/2$ & $\pm 7/5 $& $1/\sqrt{5}$  & $\sqrt{2/3}$ \\
$ \pm 3/2$ & $ \pm 3/2$ & $\pm 3/5$ & $1/\sqrt{5}$  & 1 \\
\hline
\end{tabular}
\end{center}
\label{tab_mTHz}
\end{table}%
The stability of each  line center frequency is  limited by  the long term fluctuations ($>100$~s) of the local magnetic field measured to 0.4~mG (pk-pk). It contributes an uncertainty proportional to $m_{THz}$ and of the order of 1~kHz. The short term fluctuations of the total local magnetic field are measured to 6~mG (pk-pk) and are responsible for a  $m_{THz}$-dependent broadening of the order of 20~kHz (pk-pk). The  Doppler effect on the three-photon dark lines is   415 times smaller than the Doppler effect on the optical cooling transition at 397~nm. It   broadens the three-photon dark lines by 20~kHz (FWHM) for a sample at 10~mK.  On the spectrum of figure~\ref{fig_spectre}, the measured linewidths range  from 42~kHz to 218~kHz.

\section{Demonstration of the three-photon resonance condition}
\begin{figure}[htbp]
\begin{center}
\includegraphics[height=4.5cm]{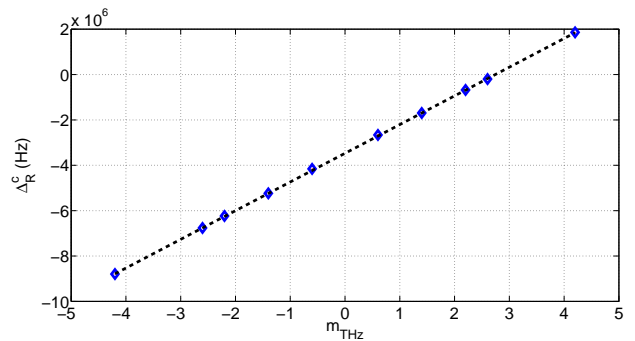}
\caption{Dark resonance frequency $\omega_R^c (m_{THz})$ of several Zeeman transition versus their corresponding $m_{THz}$ factor. The frequencies are plotted as detunings $\Delta_R^c(m_{THz})$, see text for details. The black line is the linear fit used to calculate the averaged magnetic field seen by the ions.  \label{fig_muB}}
\end{center}
\end{figure}

\begin{figure}[htbp]
\begin{center}
\includegraphics[height=4.5cm]{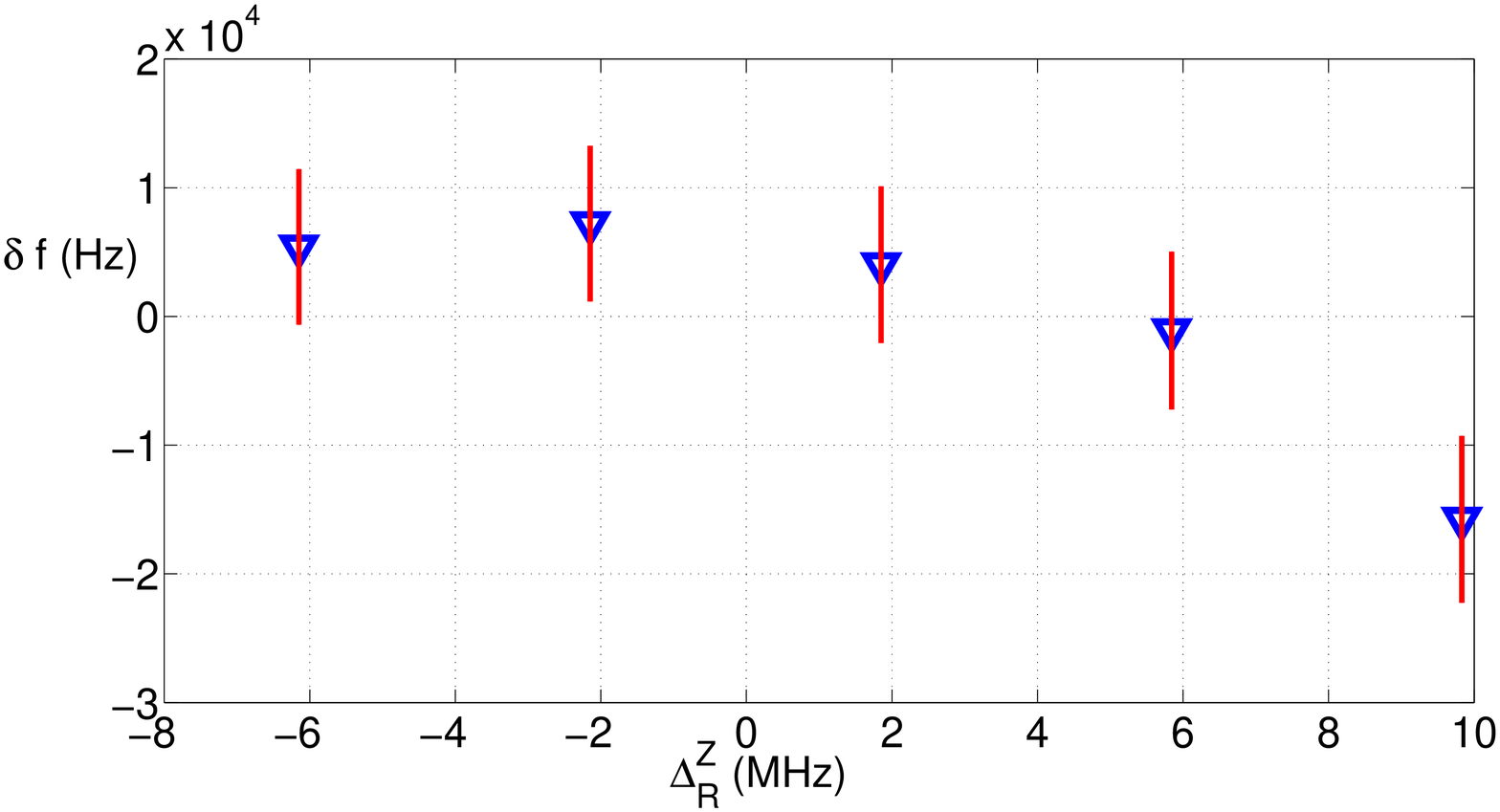}
\caption{Frequency shift $\delta f $ on the 1.82~THz transition,  for the $m_{THz}=-13/5$ dark line observed for 5 different sets of $\{\omega_C, \omega_B \}$, versus the one-photon detuning $\Delta_R^Z$  that fits the three photon resonance condition. The Zeeman effect is removed from the shift  and added to the measured one-photon detuning for the Zeeman shifted $R$-transition (Eq.~\ref{accord_D}). Fixed values : $P_B=10$~mW, $P_R=2$~mW, $P_C=8$~mW, $\Delta_B=-24.94$~MHz (errorbar$=\pm 1$ std).      \label{fig_shift3P}}
\end{center}
\end{figure}

To prove that these extra lines result from the 3-photon process and are referenced to the THz transition, we focus on the dark line defined by $m_{THz}=-13/5$ because of its high contrast. The $R$-laser frequency  $\omega_R$  is scanned  for different values of $\omega_C$ in an interval of 16~MHz, while $\omega_B$ is kept constant. The  frequency step is 1~kHz and signal is accumulated for 150~ms at each step. Each scan is reproduced 4 times and averaged. Each observed line profile is fitted to a Lorentzian profile and the center of the line $\omega_R^c$ is pointed with an uncertainty of the order of 1~kHz (1 $\sigma$) conditioned by the frequency step and the signal to noise ratio. The frequency combination $\Delta_{RCB}=\omega_R^c+\omega_C-\omega_B$ is expected to give access to the magnetic dipole transition frequency, once the experimental shifts  removed,  the Zeeman shift being the largest  one identified. As some non-negligible light-induced effects also shift the dark lines, and their dependence with the $m_{THz}$ number is not known, we do not use the average frequency of the $+m_{THz}$ and $-m_{THz}$ line to deduce the unshifted  frequency. We rather estimate the local magnetic field by exploiting several multiline spectra as the one of Fig.~\ref{fig_spectre} and adjusting a linear fit of the dark resonance frequencies $\omega_R^c (m_{THz})$ with the  $m_{THz}$ values (see Fig.~\ref{fig_muB} where they are plotted with respect to the  transition frequency $\omega_{P_{1/2}D_{3/2}}$ measured in \cite{gebert15}). 
The Zeeman shift $\delta_Z (-13/5)$ of the $m_{THz}=-13/5$ dark line is evaluated with an uncertainty of $\pm 6$~kHz dominating the   total uncertainty of the THz frequency. As shown on Fig~\ref{fig_shift3P}, the Zeeman corrected transition frequencies do not exactly match the 3D$_{3/2}$ to 3D$_{5/2}$ transition frequency $f_{DD}$  of reference \cite{solaro18} and are shifted from this reference value by amounts $\delta f =\Delta_{RCB}-\delta_Z (-13/5)-f_{DD}$ which evolve between +5 and -15 ($\pm 6)$~kHz. These shifts are 3 orders of magnitude smaller than the range covered by the one-photon detuning $\Delta_R$ and we consider that the three-photon resonance condition is demonstrated. Nevertheless, their plot against the $m_{THz}=-13/5$ transition detuning $\Delta_R^Z=\omega_R^c-\omega_{P_{1/2}D_{3/2}}+\delta_R^Z(-13/5)$,  with  $\delta_R^Z(-13/5)$ the Zeeman shift on the $R$-transition, shows a correlation between the shifts and  the detunings that cannot be explained by any drifts in the experimental set-up, neither by the coupling on the quadrupole transition $\delta_C$. Dependence of the light-induced shift with the one-photon detuning is observed in two-photon CPT for a continuous laser excitation \cite{brandt97,wynands99,zanon11} and with a Ramsey interrogation scheme \cite{hemmer89,pollock18}. When  they are not light-shifts induced by neighbour transitions, they are understood as induced by the relaxation of the coherence between the stable and metastable states involved in the dark state. This relaxation can be due to collisions but most probably to laser relative phase diffusion in our cold atom system in ultra-high vacuum (with a pressure lower than $10^{-9}$ mbar). In the case of the three-photon CPT, the only light-shifts which can be responsible for a dark line frequency shifts are the one shifting the 3D$_{3/2}$ or the 3D$_{5/2}$ Zeeman sub-states, as the other ones cancel in the three photon resonance condition (Eq.~\ref{accord_D}).

\section{Power-induced shifts}
\begin{figure}[htbp]
\begin{center}
\includegraphics[height=4.5cm]{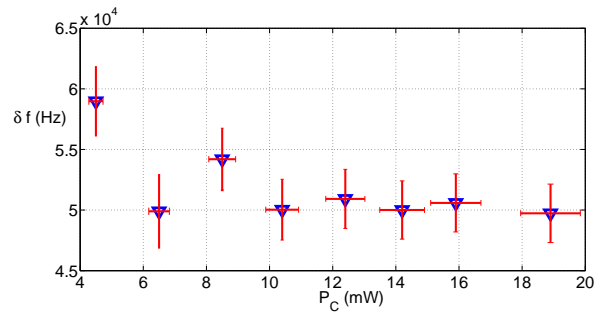}
\caption{ Frequency shift of the Zeeman sub-transition $m_{THz}=-13/5$ vs $P_C$, the laser power on the quadrupole transition, for one-photon detuning $\Delta_R^Z$=-10.6~MHz. The other laser powers are $P_B$=10~mW and $P_R$=1~mW.   \label{fig_P729}}
\end{center}
\end{figure}

 In the case of a three-photon CPT, interpreted as a laser-mediated $\Lambda$-scheme \cite{champenois06}, the relevant one-photon detuning is $\Delta_R$ and the laser couplings on the two arms of the $\Lambda$-scheme are dominated by the laser excitation on the $R$-transition. This description of the $N$-level scheme is relevant for a strong enough coupling on the weak $C$-transition. This is confirmed by results reported on Fig.~\ref{fig_P729} where, contrary to a conventional light-shift effect, it takes a minimum laser power $P_C$ for the frequency shift to reach a value independent of this power. This behaviour is attributed to  an ineffective $\Lambda$-scheme when the coupling on the quadrupole transition is too weak.  

\begin{figure}[htbp]
\begin{center}
\includegraphics[height=4.5cm]{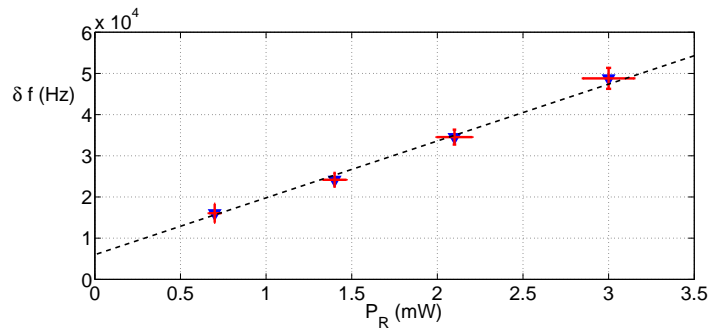}
\caption{Frequency shift of the Zeeman sub-transition $m_{THz}=-13/5$ measured for different $R$-laser power $P_R$ (blue dots) and for an effective one-photon detuning $\Delta_R^Z=-7.74$~MHz. The dark line is a linear fit of this plots, with slope $13.8\pm 1.8$~kHz/mW and limit shift for null power $P_R$ equal to  $6.0 \pm 3.4$~kHz ($1 \sigma$). The other laser powers are $P_B$=19~mW and $P_C$=10.4~mW.   \label{fig_PRshift}}
\end{center}
\end{figure}

The dependence of the frequency shift with the power of the $R$-laser  is shown on Fig.~\ref{fig_PRshift} for the $m_{THz}=-13/5$ line. The results show a  linear behavior of  $\delta f$ with $P_R$  which can be extrapolated for $P_R=0$ to $6.0(\pm 3.4)$~kHz, showing that other experimental parameters are also responsible for shifts. In the case of the $m_{THz}=-13/5$ line,  a light-shift effect can be induced by a small projection of the $R$-laser polarisation along the magnetic field axis which then couples the  3D$_{3/2}$, $m=-1/2$ sub-state to 4P$_{1/2}$, $m=-1/2$ sub-state. The light-shift induced by the far detuned $R$-couplings to the 4P$_{3/2}$ state is estimated to values lower than 10~Hz. In the regime of parameter used to observe the three-photon dark line, no significant dependence of the frequency shifts with the laser power $P_B$ could be observed, for power ranging 5 to 20~mW. Another kind of power-induced shifts have been identified in two-photon CPT-clock \cite{nagel99, zanon11}. They are due to the relaxation of the coherence between the stable and metastable states involved in the dark state  and are proportional to the one-photon detuning. This extra effect is certainly contributing to the total power-induced shift, due to the finite phase coherence of the three lasers \cite{scharnhorst15, collombon19}.

\section{metrological performances}
 To quantify the metrological performances of  the 3-photon CPT dark line as a THz reference, let's assume these shifts are under control and focus on the linewidth, the absolute signal level as well as the contrast of the dark line. We recall that within the present experimental set-up, each line is broadened by a residual  Doppler effect  (estimated to a minimum of 20~kHz FWHM) and by a fluctuating Zeeman effect (estimated to $8.4\times m_{THz}$~kHz pk-pk). Furthermore, in the range of our experimental parameters, the observed power-induced broadening   is only due to the coupling on the $R$-transition.  The narrowest observed dark lines (linewidth of 45~kHz for $P_R=0.7$~mW) are the ones with the smallest coupling on the $R$-transition, identified by $|m_{THz}|=11/5$ and 13/5 (see table~\ref{tab_mTHz}). The maximum dark line contrast, reaching 25\%, is observed for the lines $|m_{THz}|=13/5$ and 21/5 and they are the ones with the largest coupling on the $C$-transition. These two independent conditions point $|m_{THz}|=13/5$ as the transition of the best contrast/broadening compromise in the context of our present experimental set-up. 
  
\begin{figure}[htbp]
\begin{center}
\includegraphics[height=4.5cm]{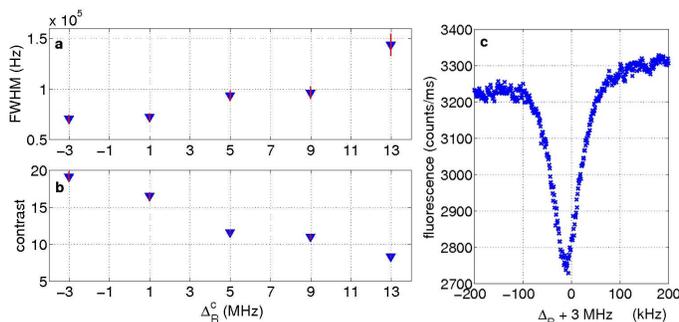}
\caption{FWHM (a) and contrast (b) of the $m_{THz}=-13/5$ dark line observed for 5 different sets of $\{\omega_C, \omega_B \}$, vs $\Delta_R^c$, same experiments as for Fig.~\ref{fig_shift3P}. (c) : line profile of the dark line showing the largest contrast and the smallest linewidth of these 5 observations. The background induced by stray light is $520 \pm 25$ counts/ms. \label{fig_FWHM}}
\end{center}
\end{figure}

In Fig.~\ref{fig_FWHM}, we have plotted  the linewidth and contrast of the $m_{THz}=-13/5$ dark line for the five different sets of $\{\omega_C, \omega_B \}$, as observed for Fig.~\ref{fig_shift3P}.  
The data show a dependence of the linewidth and contrast with the one-photon detuning $\Delta_R$ for which the CPT occurs while all the laser powers are kept constant. We think that this effect can be explained by the relative position of the dark line in the broader fluorescence spectra profiles of the trapped ions, due to a competition with the strong transitions involved in laser cooling. 
Further studies are required to identify the best condition to reach the contrast/linewidth optimum but these curves open very positive perspectives as we find in the same detuning range the smallest shifts, the largest contrast and the narrowest linewidth.

Considering that the Doppler broadening can be cancelled by obeying the phase matching condition $\vec{k}_B-\vec{k}_R-\vec{k}_C=\vec{0}$ \cite{champenois07}, or alternatively by a Lamb-Dicke effect on the THz-wavelength scale \cite{alighanbari18}, the signal over noise ratio could be increased with a larger number of ions building the dark line. By expanding the size of the cloud, the  position dependant systematic shifts are expected to broaden the spectroscopic lines    \cite{arnold15,arnold16,keller19}. Nevertheless, when varying the number of trapped ions from 60 to 400 ions, with all other parameters fixed, we observe no variation of the THz-frequency shift, neither of the dark line width, within the 6~kHz resolution.
 
As the magnetic field fluctuations can be actively reduced by a factor 50 and the sensitivity to these fluctuations can also be reduced by using the  transitions  $m_{THz}=\pm 1/5$ if other laser polarisation and propagation directions are permitted by the set-up, a linewidth in the kHz range, or lower, together with a large contrast, seems  to be very accessible. It would enable to identify effects, which are so far too small to be detected and to access a resolution in the $10^{-9}$ range, which is the state of the art in the THz precision spectroscopy \cite{hu17,alighanbari18}. Indeed, in a system without any experimentally induced decoherence, one can show that sub-kHz linewidths can be observed \cite{champenois07}.  Furthermore, with a 25\% contrast and an average fluorescence signal of 4000 counts/ms which is the typical value for the fluorescence of one thousand trapped ions in the optimum condition for a narrow dark line, the signal to noise ratio at 1~ms reaches 16. Even with a kHz linewidth, such a large signal to noise ratio   allows  the resolution to be increased to the  $10^{-11}$ range by averaging data over seconds. 

\section{Conclusion}
The demonstrated 3-photon CPT has a large potential for high-resolution spectroscopy.  Very similar to 2-photon CPT, the interrogation protocole depends on  numerous parameters that have to be further explored. In 2-photon CPT, they are strongly mitigated by the use of Ramsey-type pulsed protocol \cite{zanon15,hafiz18,abdel18} and using a pulsed interrogation method for the 3-photon CPT is certainly a route to test.  The originality of our approach allows  to access the THz domain, an insufficiently   explored and very promising  spectral domain,  so far associated to rotational transitions in light molecules. Although the production of a continuous-waveTHz radiation from the three involved optical radiations is still an  issue, this configuration could be of use to transmit the 1.82~THz reference signal by optical means and to benefit from very efficient detectors. Indeed, because THz radiation hardly propagates along long distances in air,  its phase-coherent transfer  implies to coherently duplicate its phase information onto an optical carrier\cite{nagano17}. Building a THz referenced signal from three coherent lasers avoids this duplication stage and allows its transmission over long distances through optical fibers.  Independently of the frequency range of the reference transition,   the demonstration that a  narrow line can be produced in a mesoscopic sample by implementing a Doppler-free technique  opens the route to its implementation on a large variety of atomic systems.

 This experiment has been financially supported by   the Labex FIRST-TF (ANR-10-LABX-48-01),  the A*MIDEX project (ANR-11-IDEX-0001-02), EquipEx Refimeve+ (ANR-11-EQPX-0039) all funded by the Investissements d'Avenir French Government program, managed by the French National Research Agency (ANR). Fundings from CNES (contract 151084) and R\'egion PACA are acknowledged. MC acknowledges FIRST-TF for funding and stimulating scientific environnement,  CyC acknowledges financial support from CNES and R\'egion Provence-Alpes-Cote d'Azur.  
 
\begin{appendix}

\section{Conditions for a three-photon dark resonance}\label{app}
The  condition for a three-photon dark resonance implying the weak quadrupole transition in Ca$^+$ is better explained in the dressed state picture where the system of interest is a motionless atom represented by its quantized internal states plus $n_B$  photons at 397~nm (energy $n_B \times \hbar \omega_B$), $n_R$ photons at 866~nm (energy $n_R \times \hbar \omega_R$) and $n_C$ photons at 729~nm (energy $n_C \times \hbar \omega_C$) (see Fig.1,~{\bf a} for the level scheme). The three atom-laser interactions are characterised by their Rabi frequencies  $ \Omega_B,  \Omega_R,  \Omega_C$   which are linear with the local laser electric field, whatever the nature of this interaction. The eigenvalues of the non-coupled hamiltonian  depend on  each laser detuning which are defined like $\Delta_B=\omega_B-\omega_{P_{1/2}S_{1/2}}$, $\Delta_R=\omega_R-\omega_{P_{1/2}D_{3/2}}$, and $\Delta_C=\omega_C-\omega_{D_{5/2}S_{1/2}}$ with $\omega_{P_{1/2}S_{1/2}}$, $\omega_{P_{1/2}D_{3/2}}$ and $\omega_{D_{5/2}S_{1/2}}$ the atomic transition frequencies. Like further explained in \cite{champenois06},  the dressed laser-coupled subsystem \{(S$_{1/2}, n_B, n_R, n_C$), (D$_{5/2},n_B, n_R, n_C-1$)\} can be diagonalised at the lowest order of the perturbation and the new eigenstates \{$|S\rangle, |Q\rangle$\} are then a coherent superposition of the two uncoupled states. \{$|S\rangle, |Q\rangle$\} are both coupled to the $|P\rangle$ state (P$_{1/2}, n_B-1, n_R, n_C$) through the strong dipole transition excited by the 397~nm laser, but with a very different strength, depending on the detuning $\Delta_C$ and the  Rabi frequency $\Omega_C$, which control the proportion of the two uncoupled states in the new eigenstates. Including $|D\rangle=($D$_{3/2}, n_B-1, n_R+1, n_C)$, the subsystem $\{|Q\rangle, |P\rangle, |D\rangle \}$, coupled by the two dipole transitions forms a $\Lambda$-scheme where the two feet are stable or metastable. This scheme is the paradigm of the configurations giving rise to a coherent population trapping  in a dark state when the two (meta)stable states are degenerated in the dressed state picture \cite{arimondo96}. Because the $\Lambda$-scheme was built by first diagonalising one of the interactions,  this resonance condition implies here not two but three photons and writes 
\begin{equation}
\Delta_R=\Delta_B-\Delta_C-\delta_C
\end{equation}
 where $\delta_C$ is the light-shift induced by the quadrupole coupling on the 729~nm transition \cite{champenois06}.   The laser couplings on the effective $\Lambda$-scheme are $\Omega_R$ and $\alpha_C \Omega_B$ with $\alpha_C=\Omega_C/2\Delta_C$ the mixing coefficient induced by the quadrupole coupling.
\end{appendix}


%

\end{document}